\documentstyle[prl,twocolumn,epsf,aps]{revtex}

\begin{document}
\draft
\title{Landauer-type transport theory for interacting quantum wires: Application to 
carbon nanotube Y junctions}
\author{S. Chen,$^{1}$ B. Trauzettel,$^{2}$ and R. Egger$^{1}$ } 
\address{${}^{1}$~Institut f\"ur Theoretische Physik, 
Heinrich-Heine-Universit\"at, D-40225 D\"usseldorf, Germany\\
${}^{2}$~Fakult\"at f\"ur Physik, Albert-Ludwigs-Universit\"at, D-79104 Freiburg, Germany 
}
\date{Date: \today}
\maketitle
\begin{abstract}
 We propose a Landauer-like theory for nonlinear transport in networks of 
 one-dimensional interacting quantum wires (Luttinger liquids). 
 A concrete example of current experimental focus is given by carbon nanotube
 Y junctions.  Our theory has three basic 
 ingredients that allow to explicitly solve this transport problem:
 (i) radiative boundary conditions to describe the coupling to external leads,
 (ii) the Kirchhoff node rule describing charge conservation,
 and (iii) density matching conditions at every node.  
\end{abstract}

\narrowtext

\begin{figure}
\epsfysize=5cm 
\epsffile{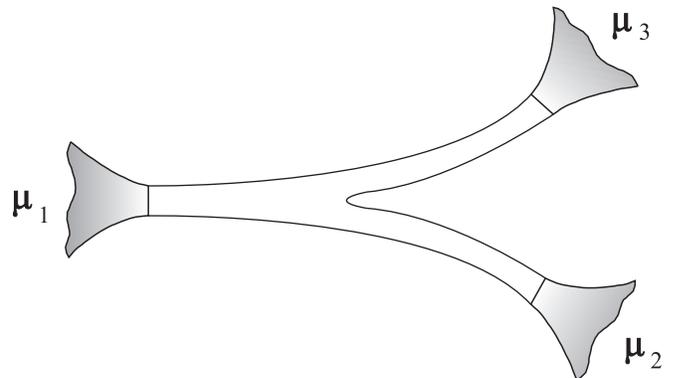} 
\caption{ \label{fig1}  Schematic view of a Y junction.  QWs
extend from adiabatically contacted 
reservoirs with electro-chemical potential $\mu_i$ at $x=-L$ to the node
at $x=0$.
}
\end{figure}

The Landauer-B\"uttiker approach to transport in mesoscopic systems 
has been very successful in describing non-interacting
electrons by using a scattering matrix formulation \cite{datta}.  
It is an important challenge to generalize
this approach to the case
of strongly correlated electrons.  Here we propose such a theory
for $N$-terminal star-like networks of interacting 1D quantum wires (QWs) described
by Luttinger liquid (LL) theory.  The two-terminal setup ($N=2$)
has been formulated and solved previously \cite{egger96,egger00}.
However, the step from $N=2$ to $N=3$, where three individually contacted
QWs meet at a single node (``Y junction''), see Fig.\ref{fig1}, 
is non-trivial yet essential
for the development of a practically useful transport theory for interacting electrons.
Recently, several works appeared where precisely this problem has been under study.  
While in Ref.~\cite{nayak} a weakly coupled ``Kondo'' node  
was considered, other authors used perturbation theory in the hopping \cite{safi}
and/or the interaction \cite{lal}. At the same time, it has become clear that
a more general approach is necessary to go beyond those special situations. 
Below we formulate boundary conditions that allow
for the explicit solution of this transport problem.
Progress along these lines is also likely to sharpen our understanding of 
conformal field theory with boundary conditions \cite{saleur}.

This problem is not only of intellectual interest but also of relevance
to transport experiments for carbon nanotubes \cite{dekker}.
As has been predicted \cite{egger97} and observed in a series
of beautiful experiments \cite{tube}, single-wall nanotubes provide a 
realization of LL physics.  Electron-electron interactions cause
 remarkably pronounced non-Fermi liquid behaviors characterized
by the standard LL parameter $g\approx 0.25$ (where $g=1$ is
the non-interacting value).  Template-based chemical vapor
deposition \cite{expy} and electron beam welding 
methods \cite{terrones} allow to fabricate and contact nanotube Y junctions,
and the intrinsic nonlinear $I-V$ characteristics of such a device has been
observed recently \cite{exp2}.  In addition, the case $N=4$ has been realized
by several groups using two crossed nanotubes \cite{crossed}, providing another
interesting application.  Furthermore, semiconductor heterostructures \cite{semi}
or ultracold trapped atomic gases \cite{petrov} may
allow for the systematic study of T- or Y-junctions as well.
Eventually it could even be possible to access the fractional statistics 
of LL quasiparticles through the noise properties of such a device,
thereby realizing a Hanbury-Brown-Twiss setup for fractionalized quasiparticles
\cite{safi}.

We study $N$ single-channel spinless QWs at $-L<x<0$
described by LL theory merging at a common node at $x=0$. For simplicity, we
assume the same interaction constant $g$ and Fermi velocity $v_F$
in each QW, with straightforward generalization also 
to include spin and flavor degrees of freedom \cite{egger97}. 
 Theory then has to address (i) the inclusion of applied voltage
sources and (ii) how a consistent treatment of the node can be achieved. 
Let us start with the first issue.
Like in the two-terminal case, adiabatically coupled 
external reservoirs held at electro-chemical potentials $\mu_i$ lead to
{\sl radiative boundary conditions} \cite{egger96} (we put $\hbar=1$)
\begin{equation}\label{bc}
\frac{ g^{-2}+1}{2} \rho_{i,R}(-L)  + 
\frac{ g^{-2}-1}{2} \rho_{i,L}(-L)  = \frac{\mu_i}{2\pi v_F} ,
\end{equation}
where $\rho_{i,R/L}(x)$ is the right/left-moving part of the density in 
QW $i$.
These boundary conditions only depend on the current injected into each
QW from the respective reservoir, which in turn does not depend on the backscattering
happening later on  at the node or within the QW.  
Here we discuss the computation of the nonlinear conductance matrix normalized to $e^2/h$,
\begin{equation}\label{coma}
G_{ij} = \frac{h}{e} \frac{\partial I_i}{\partial \mu_j} ,
\end{equation}
where the current $I_i$ flowing through QW $i$ is oriented towards the node,
leaving noise properties to a future publication.
Below  applied voltages $U_i$ are defined as
\begin{equation} \label{volt}
 eU_i = \mu_i-\bar{\mu}\;, \qquad \bar{\mu}= \frac{1}{N} \sum_i \mu_i.
\end{equation}
Under this definition, gauge invariance (conductance matrix is invariant
under a uniform change of all $\mu_i$) is automatically fulfilled if the $G_{ij}$ depend
only on the $U_i$.

Next we address the physics arising at the node $x=0$.
Conservation of charge enforces the {\sl Kirchhoff node rule} 
\begin{equation}\label{kirch}
 \sum_i I_i = 0.
\end{equation}
A second requirement at the node is the wavefunction matching via the
$S$ matrix \cite{ami,menon}
\begin{equation}\label{smat}
\Psi_L(0)= S \Psi_R(0) ,
\end{equation}
where $\Psi(x)=(\psi_1,\ldots,\psi_N)$ contains the wavefunctions for the $N$ QWs, and 
the outgoing (left-moving) components $\Psi_L$ are connected to the incoming
(right-moving) states $\Psi_R$
via an appropriate $N \times N$ matrix $S$, see, e.g.~Ref.~\cite{itoh}, for specific
choices at $N=3$.
Note that the scattering matrix in Eq.~(\ref{smat}) provides
 a ``bare'' (microscopic) description of the node
properties, while interactions could dynamically generate some different
boundary condition at low energy scales.

Unfortunately, a boundary condition like Eq.~(\ref{smat}) is very difficult to handle
for correlated electronic systems and typically does not allow for progress. 
Here we proceed differently by constructing  
a wide class of practically important $S$ matrices (albeit not 
all possible ones \cite{itoh}) in the following way.  
We first consider an {\sl ideal}\ system composed of {\sl impurity-free 
QWs symmetrically connected at the node}.
Microscopically, the corresponding non-interacting
problem could be modelled as $N$ tight-binding chains with 
equal hopping matrix element $t_0$,
where the ``last'' site of each chain is connected to the 
common node site via the same $t_0$ and all on-site energies are equal. 
Such a node corresponds to the special highly symmetric $S$ matrix
\begin{equation}\label{smatspec}
S = \left( \begin{array}{cccc} (2-N)/N & 2/N & \cdots & 2/N \\ 2/N & (2-N)/N & 
\cdots & 2/N \\ \cdots & \cdots & \cdots & \cdots \\ 2/N & 2/N & \cdots &( 2-N)/N  
\end{array} \right) .
\end{equation}
For this $S$ matrix, Eq.~(\ref{smat}) directly implies $\psi_1(0)=\cdots=\psi_N(0)$
for the components of $\Psi=\Psi_L+\Psi_R$. Since phases
in $\psi_i(0)$ can be gauged away,
a {\sl density matching condition} upon approaching the node results,
\begin{equation} \label{dens}
 \rho_1(0) = \cdots = \rho_N (0),
\end{equation} 
where $\rho_i(0)$ denotes the {\sl total}\ electronic
density in QW $i$ close to the node.
Remarkably, the conditions (\ref{bc}), (\ref{kirch}) and (\ref{dens})
then allow to explicitly solve this transport problem for arbitrary interactions
because the condition (\ref{dens}) does not involve wavefunction phases but only
amplitudes.  To arrive at more general 
$S$ matrices, in a second step we then consider additional impurities
in the different QWs displaced slightly away from the node.  Inclusion of impurities
does not cause conceptual difficulties, and such a  modelling allows
to construct all $S$ matrices of practical interest.

Let us first discuss the ideal node defined by the bare scattering matrix 
(\ref{smatspec}).  For $g=1$, the conditions (\ref{bc}), (\ref{kirch}) and 
(\ref{dens}) are in fact equivalent to the standard Landauer-B\"uttiker approach.
This is easily seen by using the usual scattering states, e.g. the state 
injected into QW $i=1$: 
\begin{equation} \label{scatstate}
\left( \begin{array}{c} \psi_1 \\ \psi_2 \\ \cdots \\ \psi_N \end{array} \right) =
\left( \begin{array}{c} e^{ikx} + r e^{-ikx} \\ t e^{-ikx} \\ \cdots \\ t e^{-ikx}
\end{array} \right)  ,
\end{equation}
with reflection (transmission) amplitude $r$ ($t$).
For $g=1$, the boundary conditions (\ref{bc}) are equivalent to a
Fermi distribution for occupying the states (\ref{scatstate}), while Eq.~(\ref{kirch}) 
gives $1=r+(N-1)t$. Furthermore, the density matching condition (\ref{dens}) yields
$|1+r|^2 = |t|^2$. Combining both equations immediately gives 
$r=(2-N)/N$ and  $t=2/N$, and hence reproduce Eq.~(\ref{smatspec}). The conductance matrix
is then 
\begin{equation}\label{nonint}
G_{ii} = 1-(1-2/N)^2 \;, \quad G_{i\neq j}= -(2/N)^2 .
\end{equation}
Will the conductance matrix  for this ideal case be affected
by interactions ($g<1$)?   Based on the two-terminal
case \cite{egger96}, one might suspect that there is no renormalization because of the 
Fermi-liquid character of the leads, and hence Eq.~(\ref{nonint})
would stay valid for arbitrary $g$.  
However, it turns out that for small applied voltages
and low temperatures, the system always flows to
the {\sl disconnected-node fixed point},
\begin{equation}\label{fixed}
G_{ij} = 0 .
\end{equation}
Thus the only {\sl stable} generic fixed point of this system for any $g<1$ 
represents an isolated node weakly connected to $N$ broken-up QWs, even
for an arbitrary $S$ matrix of the node.
The corrections to Eq.~(\ref{fixed}) due to finite $U_i$ or $T$ are then
sensitive to interactions.  This phenomenon is a 
consequence of the strong correlations in the LL, which here induce 
asymptotically vanishing currents even in a perfectly clean (impurity-free) system.
Equation (\ref{fixed}) also implies that open boundary
bosonization \cite{fabrizio} allows to access the asymptotic low-energy transport
properties of a QW network.

For arbitrary $g$ and sufficiently far away from the node such that 
Friedel oscillations \cite{friedel} have decayed and the boundary conditions
(\ref{bc}) hold, the left- and right-moving densities must combine to give
\begin{eqnarray} \label{ans1}
I_i &=& ev_F(\rho_{i,R}-\rho_{i,L})=
 \frac{e^2}{2\pi} \left(U_i-\sum_j T_{ij} U_j \right), \\ \label{ans2}
\rho_i &=& \rho_{i,R}+ \rho_{i,L} =
 \frac{e g^2}{2\pi v_F} \left (U_i + \sum_j T_{ij} U_j \right) .
\end{eqnarray}
Here the matrix $T_{ij}$ has been defined whose entries depend on $g$ and all $U_i$.
The $T_{ij}$ reduce to standard Landauer-B\"uttiker transmission 
probabilities for $g=1$, but in general cannot be interpreted as single-particle
quantities. 
It is important to stress that Eqs.~(\ref{ans1}) and (\ref{ans2})  are consistent with
both the LL equation of motion and the boundary conditions (\ref{bc}) for 
{\sl arbitrary} $T_{ij}$.
The Kirchhoff node rule can then be satisfied by requiring 
\begin{equation}\label{kirch2}
\sum_{i=1}^N T_{ij} = 1, 
\end{equation}
and we now use the density matching conditions (\ref{dens}) to obtain the $T_{ij}$
and hence the conductance matrix.   We mention in passing that
the usual relation $\sum_j T_{ij}=1$ should not be enforced since gauge invariance
under uniform potential shifts in all reservoirs has been encoded in Eq.~(\ref{volt})
already.

At this point it is crucial to realize that close to the node, the density $\rho_i(x)$
will deviate from the spatially homogeneous value $\rho_i$ in Eq.~(\ref{ans2})
due to Friedel oscillations.  This happens already for $g=1$, as can 
be easily checked
 from Eq.~(\ref{scatstate}) and the subsequent solution.
The total density very close to the node is
$\rho_i(0) = \rho_i + \delta \rho_i(0)$,
where $\delta \rho_i (0)$ is the Friedel oscillation amplitude at the node location.
The  $2k_F$ oscillatory Friedel contribution $\delta\rho_i(x)$ in QW $i$
arises due to interference of
the incoming right-movers and the left-movers that are backscattered at the node.
Importantly, left-movers that are transmitted from the other $N-1$  QWs into QW $i$
cannot interfere with the incoming right-mover and will therefore
not contribute to $\delta \rho_i(x)$.  This implies that $\delta\rho_i(x)$ is
identical to the corresponding Friedel oscillation in a two-terminal setup with
the same (bare) reflection coefficient as the one induced by the clean node, $R=(1-2/N)^2$.
Fortunately, this allows us to obtain $\delta\rho_i(0)$ in an exact manner using
the relation \cite{egger96}
\begin{equation}\label{pinning}
\delta \rho_i (0 ) =  - \frac{g^2 e V_i}{\pi v_F} ,
\end{equation}
where the ``four-terminal voltage'' parameter $V_i$ is found from a self-consistency
equation.  This equation is explicitly given and solved for arbitrary $g$ 
in Ref.~\cite{egger00},
and using this solution, we also have the pinning amplitude (\ref{pinning})
of the Friedel oscillation for any value of $g$.
The relation to the
two-terminal problem is not a necessary ingredient for our calculation, 
but quite convenient in allowing to compute  $\delta \rho_i(0)$ from known results.  

For concreteness, in what follows we consider $g=1/2$ where this self-consistency 
equation is quite simple ($k_B=1$),
\begin{equation} \label{selfcon}
\frac{eV_i}{2 T_B} = {\rm Im} \psi \left( \frac{1}{2} +
\frac{T_B +ie(U_i-V_i/2)}{2\pi T} \right)  ,
\end{equation}
where $\psi$ is the digamma function and
 $T_B=\pi\lambda^2/\omega_c$ for bandwidth $\omega_c$ and
impurity strength $\lambda$
of the corresponding two-terminal problem.  (Specifically, for $N=3$, to match the reflection
coefficient of the ideal Y junction, $(\pi\lambda/\omega_c)^2=1/(1+N/(N-2))$.  
The solution to Eq.~(\ref{selfcon}) then yields
$V_i$ as a function of $U_i$ alone, which, however, itself depends on all
the chemical potentials, see Eq.~(\ref{volt}). 
The density matching conditions (\ref{dens}) are then solved by enforcing 
\[
(1+T_{jj})U_j-2V_j = T_{kj}U_j 
\]
for all pairs $k\neq j=1,\cdots N$,
and with the Kirchhoff rule we find
\begin{eqnarray}
 T_{ii}(\{\mu_j\}) & = &\frac{2-N}{N}+ \frac{2(N-1)V_i}{N U_i}, \label{Tiig} \\
 T_{k\neq i}(\{\mu_j\}) & = &\frac{2}{N}- \frac{2V_i}{N U_i}. \label{Tijg} 
\end{eqnarray} 
Note that the $T_{ij}$ depend only on the applied voltages $U_i$ but not on the $\mu_i$.
As a result, gauge invariance is ensured.  The relations (\ref{Tiig}) and (\ref{Tijg}) 
represent the complete solution for the special $S$ matrix  (\ref{smatspec}).

\begin{figure}
\epsfysize=5cm 
\epsffile{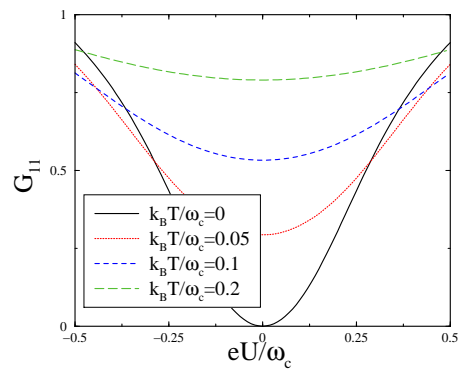} 
\caption{ \label{fig2}
 Nonlinear conductance $G_{11}(U)$ for $N=3$, $\mu_1=E_F+U$, 
$\mu_2=\mu_3=E_F$, several temperatures and $g=1/2$.
}
\end{figure}

Before proceeding to more general $S$ matrices, let us 
discuss these results for the ideal Y junction $(N=3)$ at $g=1/2$.
{}From Eq.~(\ref{coma}), we get
\begin{eqnarray*}
G_{ii} &=&  \frac{8}{9} \left(1 -  \frac{\partial  V_{i}} {\partial U_{i}} \right)
+ \frac{2}{9} \sum_{j\neq i} \left(1 -  \frac{\partial  V_{j}} {\partial U_{j}} \right)
 \\
G_{j\neq i} &=&   \frac{4}{9}  \left( \frac{\partial  V_{i}} {\partial U_i} - 1 \right)
                + \frac{4}{9}  \left( \frac{\partial  V_{j}} {\partial U_j} - 1 \right)
                -  \frac{2}{9}  \left( \frac{\partial V_{k}} {\partial U_k} - 1 \right)
\end{eqnarray*}
where $i\neq j\neq k$ in the second equation. 
Note  that the conductance matrix is symmetric.
For $e U_i \ll T$, the linear conductances follow:
\[
G_{ij} = ( 2 \delta_{ij}- 2/3) \frac{1-c\psi'(c+1/2)}{1+c\psi'(c+1/2)} ,
\]
where $c=T_B/2\pi T$ and $\psi'$ is the trigamma function. 
As $T\to 0$, the conductance matrix approaches
the stable fixed point (\ref{fixed}),  $G_{ij}\sim (T/T_B)^2$.  In general,
for $g<1$, we find $G_{ij} \sim (T/T_B)^{2/g-2}$ as $T\to 0$.  
Since this power law coincides with the prediction of
open boundary bosonization around (\ref{fixed}), this also provides a consistency
check for our calculation.
Corresponding power laws also govern the nonlinear conductances.
This is shown in  Fig.~\ref{fig2} for $g=1/2$, depicting
$G_{11}(U)$ at different temperatures.  In the $T=0$ limit, $G_{11}\sim U^2$.
Clearly, interactions have a rather spectacular effect on the
transport properties of this system.

Next we briefly outline how to construct more general node $S$ matrices
based on this solution of the ideal junction for arbitrary $g$.  The idea is 
to add impurities of strength $\lambda_i$ in each QW close to the node,
$x_i\approx -1/k_F$, which will modify the bare $S$ matrix; for explicit calculations,
see Ref.~\cite{ami}.
One can then compute the $G_{ij}$ for this more general case, but still 
allowing for arbitrary $g<1$, e.g.~by using
perturbation theory around the above solution of the ideal junction.
Focusing on $N=3$ and just one impurity,
$\lambda_2=\lambda_3=0$, to lowest order
in $\lambda_1$, a straightforward calculation gives $I_i=I_i^0 +\delta I_i$, where
$I_i^0$ is the current through QW $i$ for $\lambda_i=0$ discussed above, and
\begin{eqnarray}\label{di1} 
\delta I_1 &=&  - 2e g_+ \frac{\lambda_1^2}{\omega_c} \sin(\pi g_+)\cos(\pi g_+)
\\ \nonumber &\times& \Gamma(1-2g_+) (2\pi I_1^0/e \omega_c)^{2g_+-1} ,
\end{eqnarray}
where $g_+=4 g/3$ and $\delta I_{k\neq 1}=- \delta I_1/2$. 
Obviously, this perturbative estimate breaks down at very small energy scales 
for $g_+<1$ but is valid for all energies at $g_+>1$.   
It is straightforward to perform similar calculations for more than one impurity, other $N$,
and/or higher orders in the $\lambda_i$.  {}From Eq.~(\ref{di1}) and generalizations,
we infer that $G_{ij}=0$ is indeed the only stable fixed point.  It is also clear
that for at least one very strong impurity $\lambda_i$, the system will reduce
to one of the special cases considered previously \cite{nayak,safi,lal}.
For $g$ close to 1, we can also make explicit contact to Ref.~\cite{lal}.  
For $S$ matrices close to Eq.~(\ref{smatspec}) and weak interactions,
our solution indeed reproduces the results of Ref.\cite{lal}.
Note that the restriction to weak interactions allows to easily treat 
more general $S$ matrices \cite{lal}.

To conclude, we have proposed a Landauer-type theory for strongly 
interacting electrons in branched quantum wires such as carbon
nanotube Y junctions.  A broad class of $S$ matrices can be 
covered by formulating a suitable boundary condition (``density
matching'') to describe an ideal symmetric junction, and on top adding
effective impurities in the individual wires.  The only stable fixed
point corresponds to disconnected quantum wires, with corrections
revealing the Luttinger liquid physics via various power laws.

We thank H. Grabert and A. Komnik for useful discussions.
Support by the DFG under the Gerhard-Hess program and under Grant No.~GR 638/19
is acknowledged.

\end{document}